\documentclass[manuscript]{aastex}

\usepackage{ulem}
\usepackage{color}
\usepackage{amsmath}
\usepackage{natbib}
\title{Characterizing Intermittency of 4-Hz Quasi-periodic Oscillation in XTE J1550-564 using Hilbert-Huang Transform}

\author{Yi-Hao Su, Yi Chou, Chin-Ping Hu, Ting-Chang Yang}
\affil{Graduate Institute of Astronomy, National Central University, Jhongli 32001, Taiwan}
\email{Su: yhsu@astro.ncu.edu.tw, Chou: yichou@astro.ncu.edu.tw}

\begin{document}
\begin{abstract}
We present the time-frequency analysis results based on the Hilbert-Huang transform (HHT) for the evolution of a 4-Hz low-frequency quasi-periodic oscillation (LFQPO) around the black hole X-ray binary XTE J1550-564. The origin of LFQPOs is still debated. To understand the cause of the peak broadening, we utilized a recently developed time-frequency analysis, HHT, for tracking the evolution of the 4-Hz LFQPO from XTE J1550-564. By adaptively decomposing the $\sim4$-Hz oscillatory component from the light curve and acquiring its instantaneous frequency, the Hilbert spectrum illustrates that the LFQPO is composed of a series of intermittent oscillations appearing occasionally between 3 Hz and 5 Hz. We further characterized this intermittency by computing the confidence limits of the instantaneous amplitudes of the intermittent oscillations, and constructed both the distributions of the QPO's high and low amplitude durations, which are the time intervals with and without significant $\sim4$-Hz oscillations, respectively. The mean high amplitude duration is 1.45 s and 90\% of the oscillation segments have lifetimes below 3.1 s. The mean low amplitude duration is 0.42 s and 90\% of these segments are shorter than 0.73 s. In addition, these intermittent oscillations exhibit a correlation between the oscillation's rms amplitude and mean count rate. This correlation could be analogous to the linear rms-flux relation found in the 4-Hz LFQPO through Fourier analysis. We conclude that the LFQPO peak in the power spectrum is broadened owing to intermittent oscillations with varying frequencies, which could be explained by using the Lense-Thirring precession model.
\end{abstract}

\keywords{accretion, accretion disks --– black hole physics --- methods: data analysis ---  X-rays: binaries –-- X-rays: individual (XTE J1550-564)}

\section{Introduction}
Low-frequency quasi-periodic oscillations (LFQPOs) with frequencies ranging from a few millihertz to $\sim20$ Hz have been detected in many black hole X-ray binaries during specific X-ray outburst states \citep{Belloni2010}. These oscillations are appropriately named to reflect their finite-width peaks, usually described by multi-Lorentzian components \citep{Belloni2002, Rao2010}, in the Fourier power spectra of their X-ray light curves. \citet{Remillard2002} discovered $\sim70$ LFQPOs in the X-ray nova XTE J1550-564 during its 1998-1999 outburst, analyzing the observations of the Proportional Counter Array (PCA) on board the Rossi X-ray Timing Explorer (RXTE). The optical counterpart identification and the dynamical measurements of XTE J1550-564 confirmed that the object is an X-ray binary consisting of a $9.10 \pm 0.61 M_{\odot}$ black hole and a $0.30 \pm 0.07 M_{\odot}$ low-mass star at a distance of $4.38^{+0.58}_{-0.41}$ kpc \citep{Orosz2011}. In addition, \citet{Steiner2011} measured the spin of the black hole in XTE J1550-564 by modeling both the thermal continuum spectrum and the Fe K$\alpha$ line, and obtained a moderate spin value of $0.49^{+0.13}_{-0.20}$.

The LFQPO's broad peak implies that the corresponding X-ray light curve is not strictly periodic and could be caused by a periodic oscillation with finite coherence time, an oscillation with varying frequency, or multiple oscillations with different frequencies. Using time-frequency analysis techniques to track the evolution of LFQPO within an observation can help to understand the origin of peak broadening. Owing to the numerous observations of LFQPOs in XTE J1550-564, this black hole X-ray binary is a good target for studying the time-frequency properties of LFQPOs. \citet{Lachowicz2010} selected a strong 4-Hz LFQPO (MJD 51085) among the LFQPOs observed from XTE J1550-564 and calculated the spectrogram, which is a time-frequency distribution of power spectra produced by dividing a light curve into small segments and transforming these segments into a sequence of Fourier spectra or Lomb-Scargle periodograms along the time axis, as the first step toward tracking the LFQPO's evolution. However, because of the conflict between the time and frequency resolutions of the spectrogram, the method was not deemed suitable for additional analysis. To overcome the time-frequency resolution limitation of the spectrogram, \citet{Lachowicz2010} used advanced time-frequency analysis methods, the wavelet analysis and the Matching Pursuit algorithm, to analyze the 4-Hz LFQPO. They found that the LFQPO consists of multiple independent oscillations with lifetimes of $\sim3$ s. In spite of the improved time-frequency resolution, these time-frequency analysis methods may suffer from issues related to understanding the underlying physics, because the mathematical assumptions of these methods are too strict to be consistent with the properties of natural signals \citep{Wu2011}. In addition, \citet{Uttley2005} showed that the X-ray variability processes of black hole X-ray binaries are non-linear. Therefore, to understand the nature of LFQPOs, we need to adopt a method which is applicable not only to non-stationary but also to non-linear processes.

The Hilbert-Huang transform (HHT) is an alternative method that could be useful for tracking the evolution of LFQPOs. \citet{Huang1998} developed this time-frequency analysis method to decompose a signal, that could be generated by non-stationary and non-linear processes, into basis components defined by the signal itself (the adaptive decomposition), and transform these components into instantaneous frequencies as functions of time. As a result, the HHT does not make strictly mathematical assumptions regarding the oscillatory components, while at the same time it has a satisfactory time-frequency resolution. Consequently, this method has found many applications in different fields \citep{Huang2008}. Specifically, the HHT 
method has been recently successfully applied to analyzing the QPO from the active galactic nucleus RE J1034+396  \citep{Hu2014}.

Owing to the progress of HHT since its introduction \citep{Huang2008, Wang2014} and owing to the successful application of this method to the QPO of RE J1034+396, HHT could be suitable for revealing the origin of the LFQPO peak broadening in the power spectrum of the black hole X-ray binary. Therefore, we employed the HHT method to study the detailed time-frequency variation of the 4-Hz LFQPO of XTE J1550-564.

In this paper, we present our HHT-based analysis of the time-frequency properties of the 4-Hz LFQPO in XTE J1550-564. In section \ref{obs}, we briefly describe the observation of this LFQPO and the data reduction process. In section \ref{analysis}, we explain how to use the HHT for adaptive decomposition of the $\sim4$-Hz oscillatory component from the X-ray light curve; we then present how to acquire the instantaneous frequency and amplitude with confidence limits and how to characterize the intermittent $\sim4$-Hz oscillations. The interpretation of the intermittency is provided in section \ref{interpretation}. Finally, we end with our conclusion in section \ref{conclusion}.

\section{Observation}\label{obs}
The 4-Hz LFQPO in XTE J1550-564 was detected by the RXTE/PCA observation on 1998 September 29 (ObsID: 30191-01-15-00) \citep{Remillard2002}. Following the data reduction process of \citet{Lachowicz2010}, we used the data in the first part of the PCA observation, which were generated from single-bit mode data with 125-$\mu$s resolution, for extracting a 2505-s-long light curve in 2-13 keV during the operation of all five Proportional Counting Units. All photon arrival times were first corrected to the barycenter of the solar system, following by that the X-ray photons were evenly binned into a light curve (bin size of 0.01 s) for subsequent HHT analysis.

\section{Hilbert-Huang Transform Analysis}\label{analysis}
\begin{figure}
\plotone{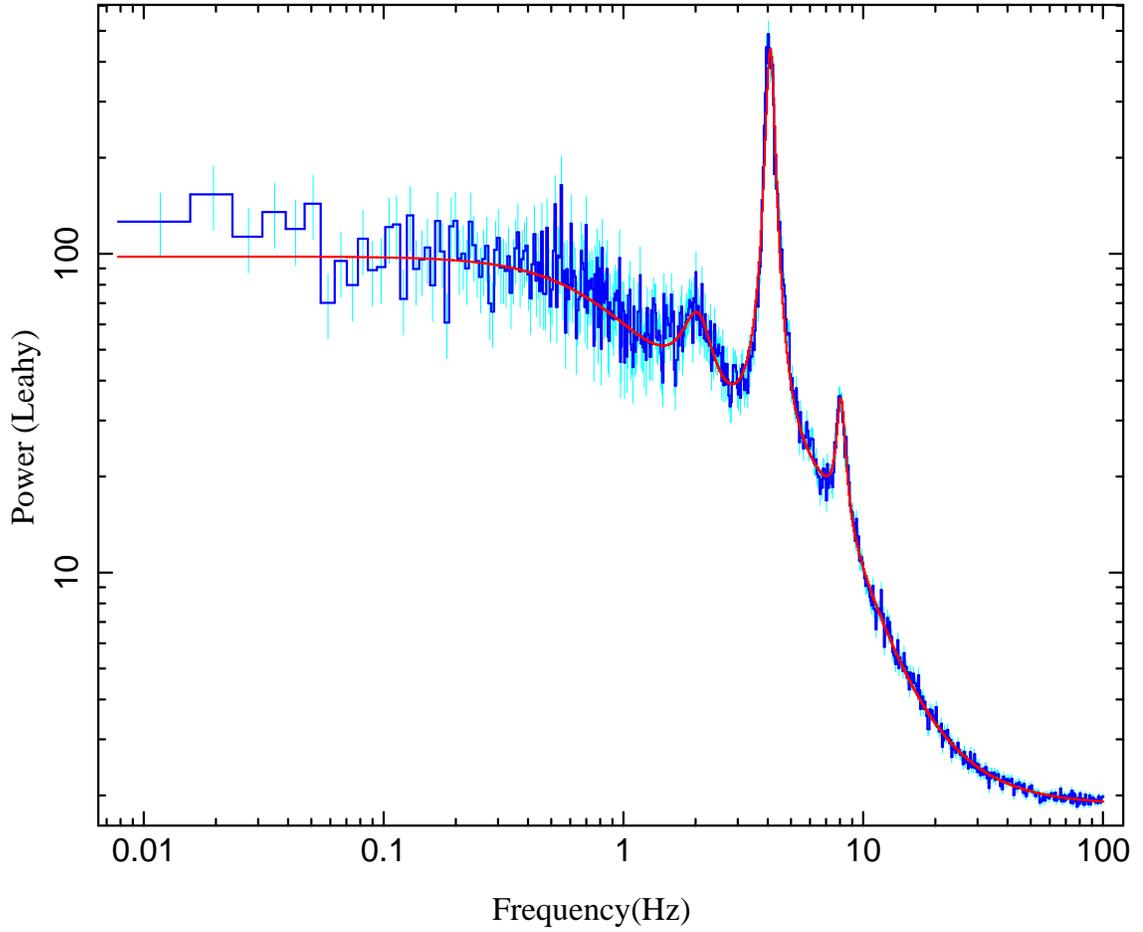} \caption{The Fourier power spectrum of the 4-Hz LFQPO in the black hole X-ray binary XTE J1550-564 (blue), plotted along with the best multi-Lorentzian model fit (red). \label{power_spectrum}}
\end{figure}

First, we briefly examined the LFQPO peak in the Fourier power spectrum by using the standard procedure for detecting the LFQPO features (Rao et al. 2010). The SITAR\footnote{http://space.mit.edu/cxc/analysis/SITAR/distrib.html} timing analysis package of the ISIS software\footnote{http://space.mit.edu/asc/isis} \citep{Houck2000} was applied to yield a set of 128-s-long power spectra and to fit the average power spectrum with Lorentzian components, revealing its finite-width peak at $4.10 \pm 0.01$ Hz and a $10.9 \pm 0.7$ quality factor, a ratio of the peak frequency to Lorentian FWHM (Figure \ref{power_spectrum}). After confirming the 4-Hz LFQPO peak, we proceeded by performing the time-frequency analysis via HHT. We utilized the HHT package of MATLAB\footnote{http://rcada.ncu.edu.tw/research1.htm} 
to track the evolution of the 4-Hz QPO from XTE J1550-564. 

The HHT algorithm consists of two main steps \citep{Huang2008}: (1) adaptively decomposing a time series into intrinsic mode functions (IMFs) by an empirical mode decomposition (EMD) method, and (2) obtaining the instantaneous frequencies and amplitudes of the IMFs through the Hilbert transform. The adaptive decomposition means that the basis components are not base on any strictly mathematical form and are derived by the signal itself. In contrast, the Fourier and wavelet analysis decompose a signal base on trigonometric and wavelet functions, respectively.  The EMD basis components, IMFs, are locally amplitude-frequency modulated time series directly extracted from the signal (see section \ref{CEEMD} for the details) and can be Hilbert transformed for accurately obtaining the instantaneous frequency and amplitude (see section \ref{HSA} for the details). Unlike the Fourier and wavelet analysis that assume a constant frequency or a waveform within a given time interval, with the temporal locality of IMFs, HHT provides a careful treatment for evolving physical systems \citep{Wu2011}. The detailed comparison among Fourier transform, wavelet and HHT are listed in the Table 1 of \citet{Huang2008}.

\subsection{Adaptive Decomposition of the X-ray Light Curve}\label{CEEMD}
For efficient adaptive decomposition of the $\sim4$-Hz oscillatory component, we adopted the fast complementary ensemble empirical mode decomposition (CEEMD) method with post-processing \citep{Wu2009, Yeh2010, Wang2014}, which is the most recently developed modified version of EMD. EMD is an adaptive decomposition method that separates oscillatory modes from original time series by iteratively subtracting the local means of local extrema in the data \citep{Huang1998, Huang2008}. These oscillatory modes at different time scales are IMFs and were defined by \citet{Huang1998}, subject to the following conditions: (1) in the entire data set, the number of extrema and the number of zero crossings must either be equal or differ at most by one, and (2) at any point, the mean value of the envelope defined by the local maxima and the envelope defined by the local minima is zero. To satisfy these conditions, the EMD is implemented by the sifting process through (1) identifying all the local extrema of the original time series and separately connecting all the local maxima and minima with cubic spline lines to form the upper and lower envelopes, (2) subtracting the mean of the envelopes from the original time series and denoting the result as a proto-IMF, (3) repeating the above two steps on the proto-IMF several times until the proto-IMF meets the conditions and denoting the result as the first IMF, (4) repeating the above steps to the residual to iteratively extract all the IMFs from the original time series until the final residual contains no more than one extremum. Therefore, the original time series, $x(t)$, can be expressed as the sum of IMFs, $c_j(t)$, and the final residual, $r_n(t)$:
\begin{equation}
x(t)=\sum_{j=1}^{n}c_j(t)+r_n(t).
\end{equation}
The detailed illustration of sifting process can be seen in the Figure 2 of \citet{Huang2008}.

If the original time series contains intermittent processes, the EMD may suffer from having oscillations with similar timescales residing in different IMF components, leading to mode mixing. To avoid this mode mixing, an improved version of the EMD, the CEEMD, was proposed \citep{Wu2009, Yeh2010}. By (1) adding a white noise series to the original time series, (2) decomposing the noisy data into IMFs via EMD, (3) repeating the above two steps several times but with different white noise series, and (4) taking the ensemble averages of these IMFs to eliminate the effect of adding the noise, the CEEMD can overcome the mode mixing problem because the added white noise provides uniformly distributed references of different timescales so that a noisy oscillation with similar timescale could be found in the corresponding IMF component \citep{Wu2009, Yeh2010}. Although the ensemble averages of the IMFs are not necessarily the IMFs, the conditions on IMFs can still be satisfied by applying the EMD post-processing to the ensemble averages of the IMFs \citep{Wu2009}. We encountered the mode mixing problem when we used EMD. Therefore, to overcome the problem and to increase the computational efficiency, we employed the fast algorithm for CEEMD \citep{Wang2014} with post-processing, for decomposing the 2505-s-long light curve.

\begin{figure}
\plotone{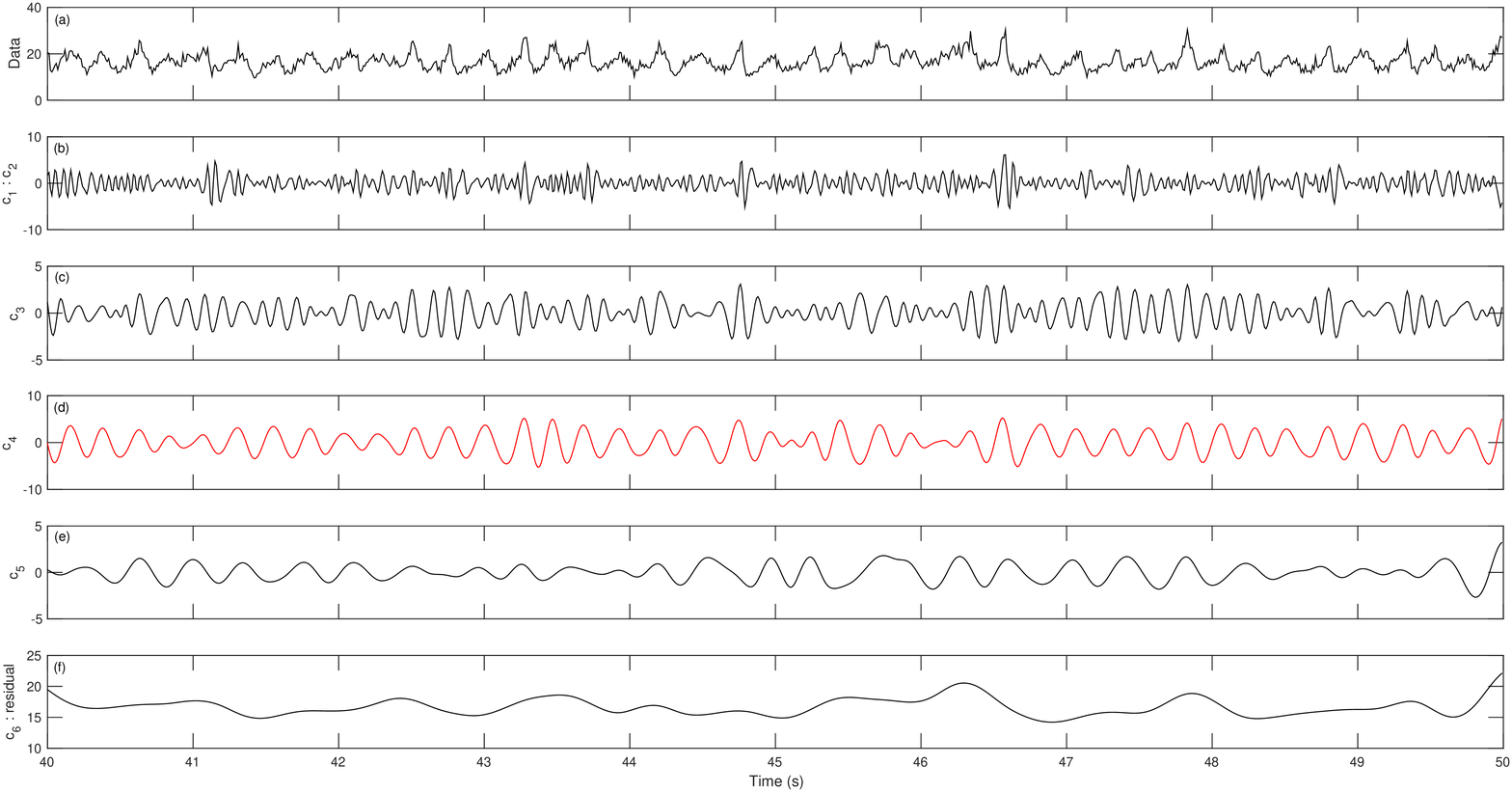} \caption{A typical example of a 10-s-long light curve and the corresponding IMFs. (a) The original light curve. (b) The high-frequency noise, which is the summation of IMF $c_1$ to $c_2$. (c) Third component, $c_3$, corresponding to the LFQPO's harmonic. (d) Fourth component, $c_4$, corresponding to the $\sim4$-Hz oscillation. (e) Fifth component, $c_5$, corresponding to the LFQPO's sub-harmonic. (f) The low-frequency noise, which is the summation of IMF $c_6$ to the final residual. On the y axis, the units are $10^{3}$ counts s$^{-1}$. \label{imfs}}
\end{figure}

\begin{figure}
\plotone{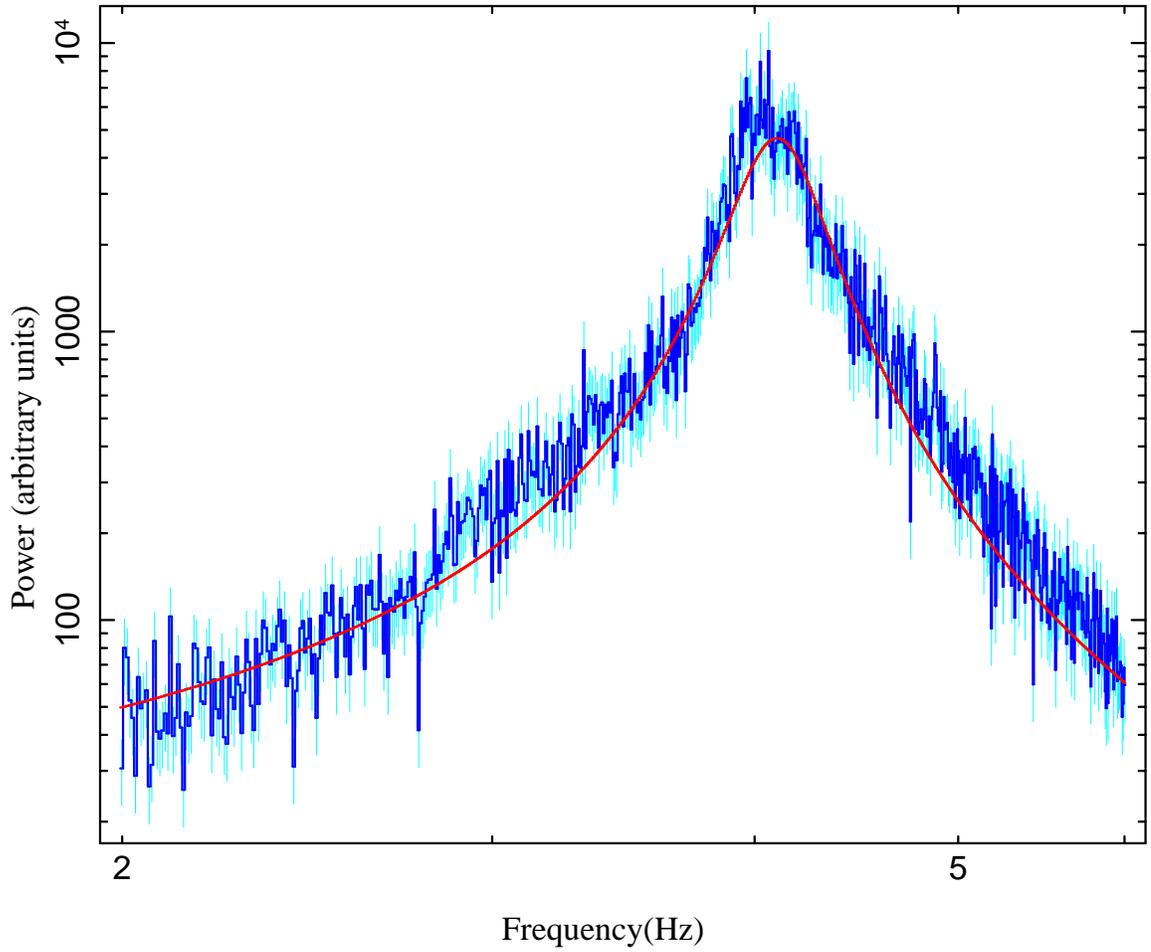} \caption{The Fourier power spectrum of the IMF $c_4$ (blue), plotted along with the best Lorentzian model fit (red). \label{power_spectrum_imf4}}
\end{figure}

Figure \ref{imfs}(a) shows a typical example of a 10-s-long light curve with a $\sim4$-Hz oscillation that can be roughly seen by directly inspecting the light curve. After decomposing the light curve and confirming the orthogonality of the IMF components \citep{Huang1998, Hu2014}, we found three significant IMF components, denoted as $c_3$, $c_4$ and $c_5$; these components are shown in Figures \ref{imfs}(c), (d) and (e). A clear $\sim4$-Hz oscillation can be observed in IMF $c_4$, which the 4 means the fourth component instead of the 4 Hz, and the oscillations in IMF $c_3$ and IMF $c_5$ are its harmonics. The high-frequency noise (summation from $c_1$ to $c_2$) and the low-frequency noise (summation from $c_6$ to the final residual) are also plotted in Figures \ref{imfs}(b) and (f), respectively. Using the adaptive decomposition, these zero-mean oscillatory components can yield the physically meaningful instantaneous frequency obtained by using the Hilbert transform \citep{Huang1998, Huang2008}. We only focused on the IMF $c_4$ in this study because it corresponds to the 4-Hz LFQPO. Figure \ref{power_spectrum_imf4} shows the average Fourier power spectrum of the IMF $c_4$ fit with a Lorentzian component. Its finite-width peak is $4.099 \pm 0.007$ Hz with a quality factor of $9.41 \pm 0.35$. This fitting result is consistent with the result from the original light curve.

\begin{figure}
\plotone{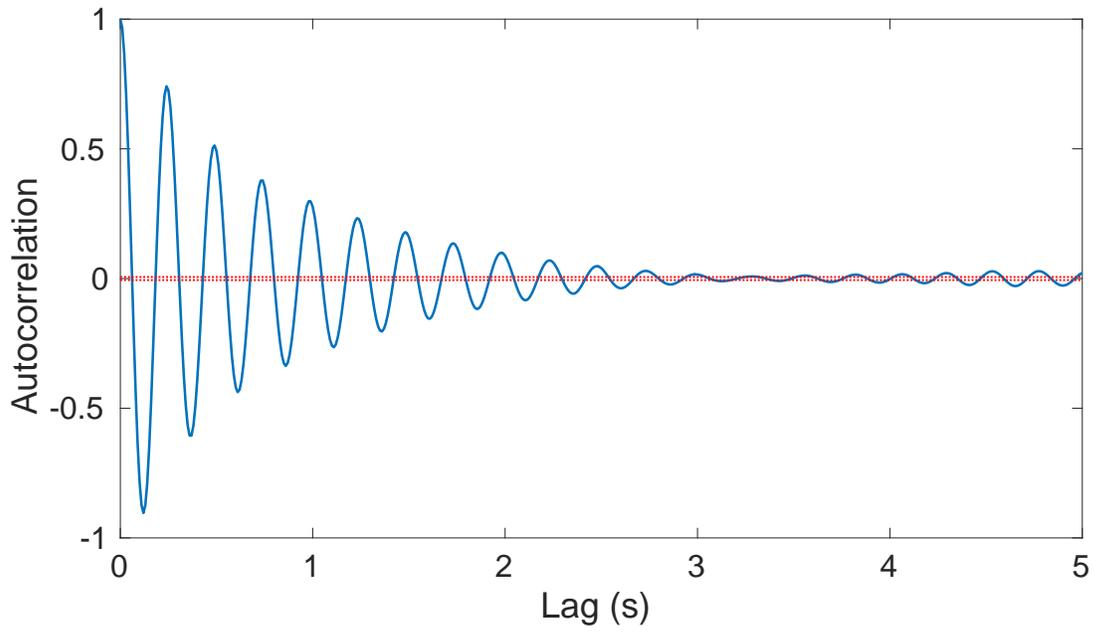} \caption{The autocorrelation function of the IMF $c_4$ with 3-sigma confidence bounds (red dotted lines). The decay of the autocorrelation function indicates that on average the duration of the $\sim4$-Hz oscillations is below 3 s. \label{autocorr}}
\end{figure}

To quickly examine the proposition that the LFQPO consists of intermittent oscillations with lifetimes of $\sim3$ s \citep{Lachowicz2010}, we calculated the autocorrelation function of the IMF $c_4$ with 3-sigma confidence bounds (red dotted lines), and the results are shown in Figure \ref{autocorr}. The decaying autocorrelation suggests that the average lifetime of the $\sim4$-Hz oscillations is shorter than 3 s, which is consistent with the result of \citet{Lachowicz2010}. This autocorrelation is also consistent with the fitting result of the quality factor, Q $\sim 10$, because there are $\sim$ 10 cycles before the oscillation decays away. Intermittent oscillations will be analyzed in more details in section \ref{characterization} via additional HHT analysis.

\subsection{Instantaneous Frequency and Amplitude of the $\sim4$-Hz Oscillation}\label{HSA}
After adaptively decomposing the $\sim4$-Hz oscillatory component, IMF $c_4$, from the original X-ray light curve, we proceeded to the second step of the HHT algorithm: using the normalized Hilbert transform \citep{Huang2009} to acquire the instantaneous frequency and amplitude of the IMF $c_4(t)$. The normalized Hilbert transform was proposed to overcome the limitation of the Bedrosian theorem, which can lead the calculation of the instantaneous frequency to be incorrect when the amplitude modulation is not very slow in comparison to the frequency modulation \citep{Huang2009}, so that one can accurately obtain the instantaneous frequency by first separating the IMF into amplitude modulation and frequency modulation parts, and then Hilbert transforming the frequency modulation part alone. Following \citet{Huang2009} and \citet{Hu2014}, we defined the instantaneous amplitude of the IMF as the cubic Hermite spline envelope of the local maxima of the absolute values of the IMF $c_4(t)$ and denoted it as $a_4(t)$. The Hilbert transform of the normalized IMF, which is the frequency modulation part alone, $X_4(t)=c_4(t)/a_4(t)$ can then be represented as
\begin{equation}
Y_4(t)=\frac{1}{\pi}P\int_{-\infty}^{\infty}\frac{X_4(t')}{t-t'}dt'
\end{equation}
where P is the Cauchy principal value of the complex integral\footnote{In practical application, instead of directly integrating the equation (2), the MATLAB accomplished the Hilbert transform by using the equation $Z_4(t)=\mathcal{F}^{-1}\{2U\{\mathcal{F}[X_4(t)]\}\}=X_4(t)+iY_4(t)$, where $\mathcal{F}$ and $\mathcal{F}^{-1}$ are the Fourier transform  and its inverse transform, respectively, $Y_4(t)$ is the Hilbert transform of $X_4(t)$, and U is the unit step function ($f>0$, $U=1$; $f<0$, $U=0$ and $f=0$, $U=1/2$, where f are corresponding Fourier frequencies). The details of the analytic signal and the relationship between Hilbert transform and Fourier transform can be seen in these links: http://www.mathworks.com/help/signal/ref/hilbert.html, http://docs.scipy.org/doc/scipy-0.15.1/reference/generated/scipy.signal.hilbert.html.}. 
Furthermore, we can define an analytical signal $Z_4(t)$ as well as the instantaneous phase function $\theta_4(t)$, which is an evenly spaced time series like $X_4(t)$ and $Y_4(t)$, as
\begin{equation}
Z_4(t)=X_4(t)+iY_4(t),
\end{equation}
\begin{equation}
\theta_4(t)=\tan^{-1} \frac{Y_4(t)}{X_4(t)}.
\end{equation}

Consequently, the instantaneous frequency of the IMF $c_4(t)$ is
\begin{equation}
\nu_4(t)=\frac{1}{2\pi}\frac{d\theta_4(t)}{dt}.
\end{equation}

\begin{figure}
\plotone{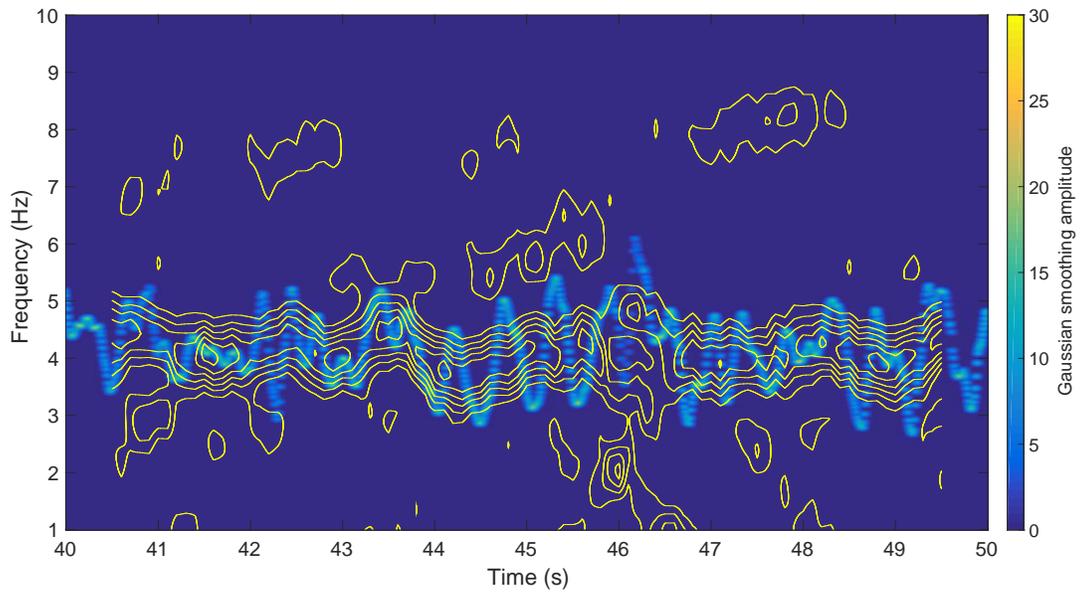} \caption{The resultant Hilbert spectrum of the 4-Hz LFQPO from XTE J1550-564. The amplitude is smoothed by a Gaussian filter for clarity. The contour plot is the Lomb-Scargle spectrogram, presented for comparison. \label{Hilbert_spectrum}}
\end{figure}

In contrast to the the Fourier and wavelet analysis, the calculation of the frequency in HHT is a differentiation over local time domain. Therefore, we can obtain the instantaneous frequency of a signal as long as the sampling time interval, $dt$ , is much shorter than the cycle length. A typical resultant instantaneous frequency and amplitude are shown as the color map for the Hilbert spectrum in Figure \ref{Hilbert_spectrum}, calculated for the time interval from 40 to 50 s. The color depth represents the magnitude of the amplitude, smoothed by a Gaussian filter for clarity, allowing to simultaneously track the frequency and amplitude variations of the $\sim4$-Hz oscillatory component. For comparison, we also show the contour plot of the calculated Lomb-Scargle spectrogram in Figure \ref{Hilbert_spectrum}. The window size and the moving step of the spectrogram were set to 1 s and 0.1 s, respectively. By comparing the Hilbert spectrum and the spectrogram, it is evident that the LFQPO's frequency is changing with time between 3 Hz and 5 Hz. However, the spectrogram's resolution precludes further analysis. By contrast, the Hilbert spectrum has a better time-frequency resolution and reveals more details regarding the $\sim4$-Hz oscillatory component, providing the strong evidence that this LFQPO shows the frequency varying and intermittent phenomenon.  

\subsection{Confidence Limits of the Instantaneous Frequency and Amplitude}\label{confidence}
As mentioned in section \ref{CEEMD}, we solved the mode mixing problem by adding white noise series to the original time series. Although the added white noise series cancel out after computing the ensemble mean, the results would be slightly different when the CEEMD is repeated with different sets of white noise series. However, we can take advantage of this property for obtaining the confidence limits for the IMF $c_4$, its instantaneous frequency, and amplitude \citep{Huang2003, Huang2008, Wu2009}. To determine the confidence limits, we first generated one thousand different sets of IMF $c_4$ by repeating the CEEMD with different sets of white noise series, and then calculated the means and the standard deviations of the IMF sets. We also calculated the means and the standard deviations of the instantaneous frequency and amplitude sets after Hilbert transforming these IMF sets.

\begin{figure}
\plotone{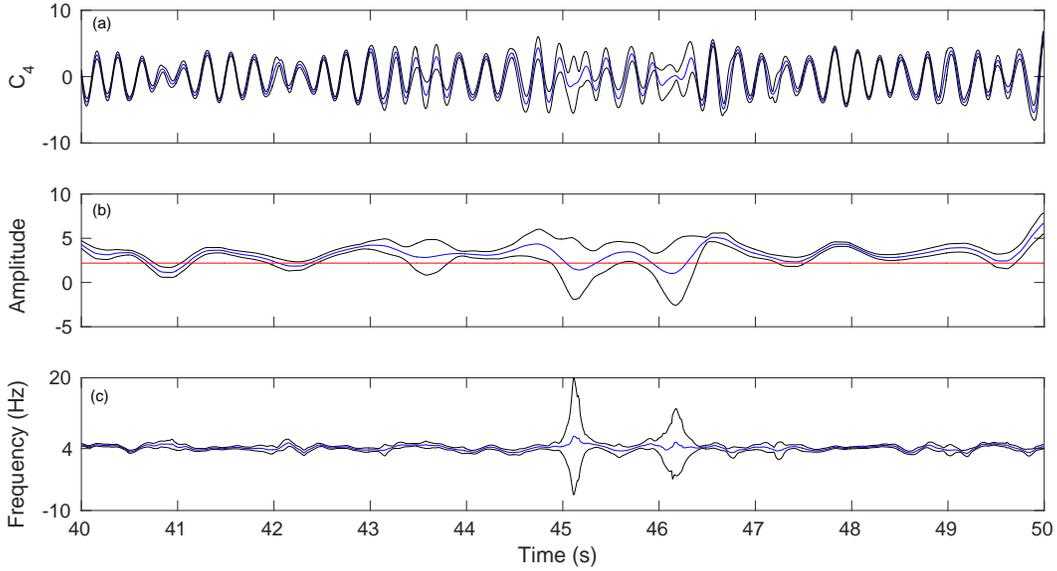} \caption{Confidence limits for the (a) IMF $c_4$, (b) instantaneous amplitude and (c) instantaneous frequency. The blue and the black lines are the means and their 3-sigma confidence limits, respectively. The red line is the average of the 3-sigma lower limit of the instantaneous amplitude. This red line is the threshold for distinguishing between the high and low amplitude durations of the intermittent oscillations. In (a) and (b) , the units on the y axis are $10^{3}$ counts s$^{-1}$. \label{confidence_limit}}
\end{figure}

Figure \ref{confidence_limit} shows the confidence limits for the IMF $c_4$ (a), its instantaneous amplitude (b) and frequency (c). The blue and the black lines are the means and their 3-sigma confidence limits, respectively. In addition, we defined the low amplitude intervals as the intervals in which the corresponding mean amplitudes (blue line in Figure \ref{confidence_limit}(b)) are below the average of the 3-sigma lower limit (red line) where no significant signal has been detected. The high amplitude durations are classified when the corresponding mean amplitudes are above the red line. These durations will be more constrained with the improved quality data and less constrained with more noisy data. Note that the largest uncertainties of IMF $c_4$, amplitude, and frequency are contributed by the low amplitude intervals.

\subsection{Characterization of the Intermittent Oscillations}\label{characterization}
\begin{figure}
\plotone{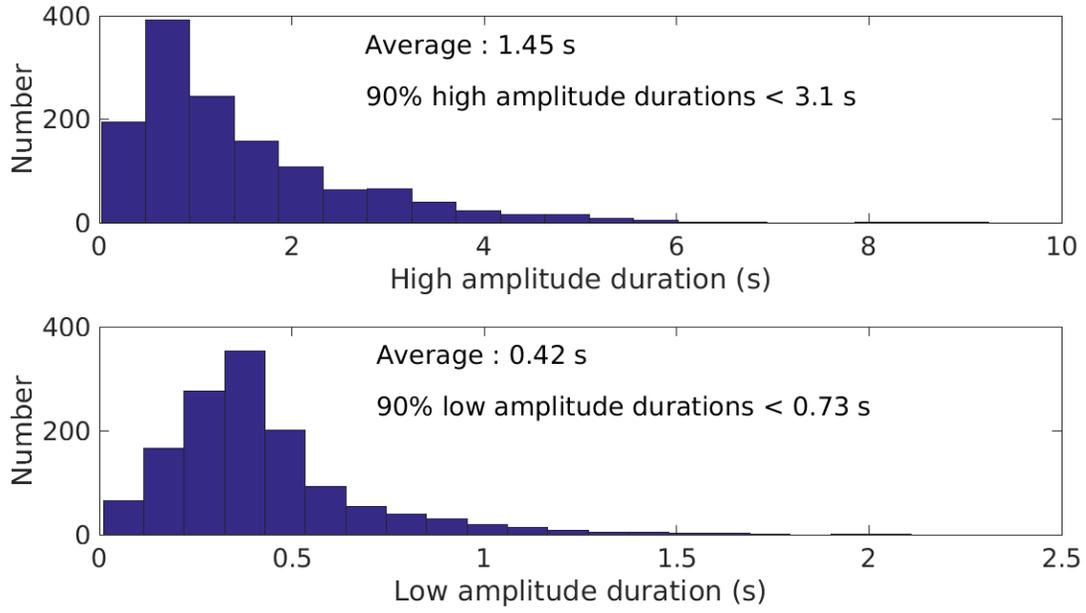} \caption{The distributions of the high amplitude (top) and low amplitude (bottom) durations. \label{lifetime_distribution}}
\end{figure}

After analyzing the 4-Hz LFQPO via HHT, we found the frequency varying and intermittent phenomenon in this LFQPO. Moreover, we utilized the confidence limit of the IMF $c_4$ instantaneous amplitude and set a threshold for determining the time interval has a significant $\sim4$-Hz oscillation, the high amplitude duration, yielding 1339 individual oscillations for the entire period of observation. We further found the distributions of the high and low amplitude durations (Figure \ref{lifetime_distribution}). The mean high amplitude duration is 1.45 s and 90\% of the oscillation segments have lifetimes below 3.1 s, which is consistent with the autocorrelation function of IMF $c_4$ (Figure \ref{autocorr}). The mean low amplitude duration, the time interval without a significant $\sim4$-Hz oscillation, is 0.42 s and 90\% of these intervals are shorter than 0.73 s.

\begin{figure}
\plotone{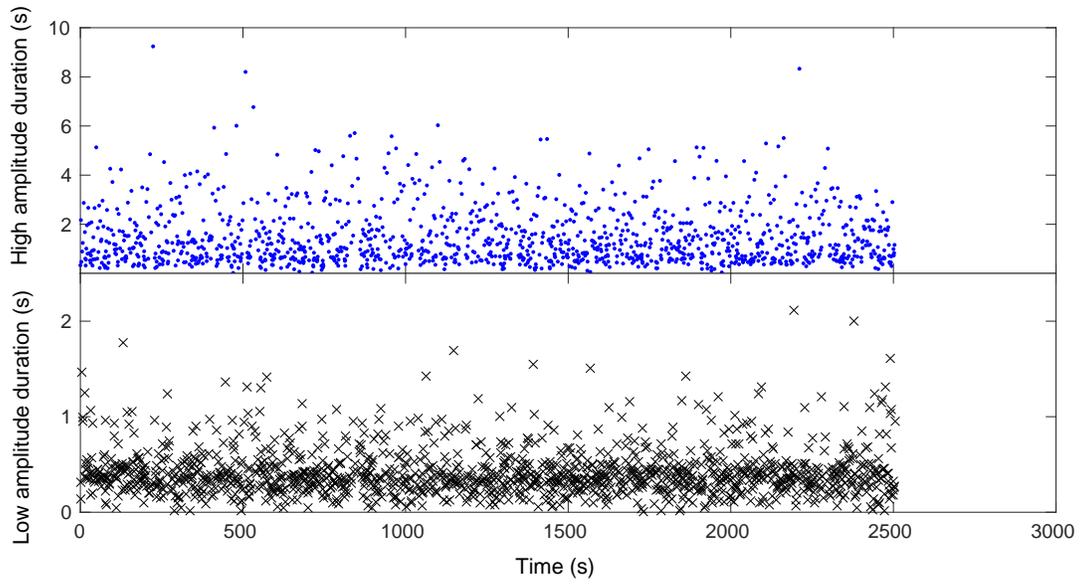} \caption{Temporal distributions of the high amplitude (top) and low amplitude (bottom) durations during the entire period of observation. \label{lifetime_time}}
\end{figure}

Figure \ref{lifetime_time} shows the temporal distributions of the high and low amplitude durations during the entire period of observation. The intervals are distributed almost uniformly, indicating that this intermittent phenomenon persists throughout the observation. A detailed discussion on the origin of these $\sim4$-Hz intermittent oscillations is given in section \ref{interpretation}.

\begin{figure}
\plotone{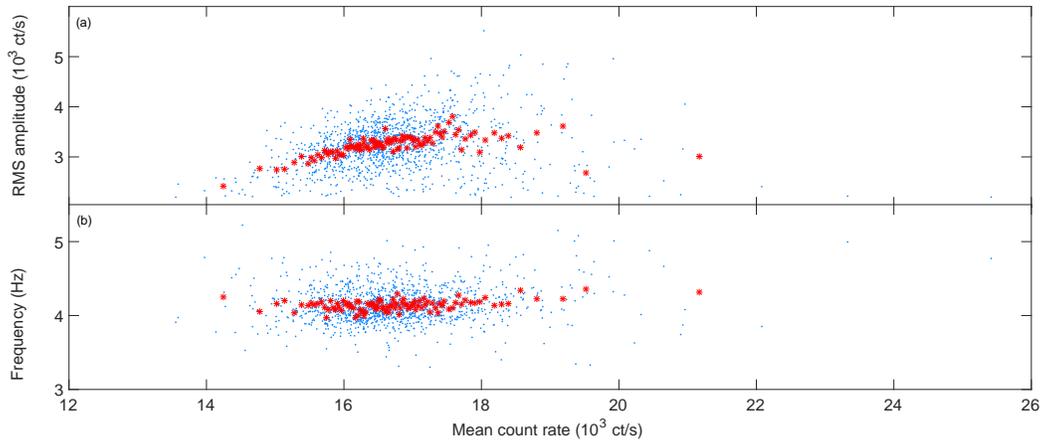} \caption{(a) Rms amplitude-flux relation for the $\sim4$-Hz intermittent oscillations. (b) Frequency-flux relation for the $\sim4$-Hz intermittent oscillations. The correlation coefficients for non-rebinned points (blue dots) are 0.213 and 0.13, respectively. After binning on mean count rate (red stars), the linear relationships become more clear and  their correlation coefficients are 0.51 and 0.41, respectively. \label{rms-flux}}
\end{figure}

\citet{Heil2011} analyzed 48 LFQPOs observed from XTE J1550-564 and demonstrated that their rms amplitude and frequency are both correlated with the source flux on short time-scales ($\sim3$ s). Because HHT analysis yields better frequency and amplitude resolutions, we obtained these relations \citep{Uttley2001, Uttley2005} from our analysis results by calculating the correlations between the rms amplitude, the mean frequency, and the mean count rate, for each high amplitude duration. Figure \ref{rms-flux}(a) shows the rms-flux relation with a calculated 0.213 correlation coefficient and a $3.3\times10^{-15}$ null hypothesis probability, indicating a strong correlation between the rms amplitude and the mean flux. The frequency-flux relation is shown in Figure \ref{rms-flux}(b) and the calculated correlation coefficient is 0.13 with a $1.8\times10^{-6}$ null hypothesis probability. This HHT-derived result is indeed similar to the result obtained from Fourier analysis by \citet{Heil2011}.

\section{Interpretation of the Intermittent Oscillations: Lense-Thirring Precession}\label{interpretation}
Our HHT-based time-frequency analysis results demonstrate that the broad spectral peak of the 4-Hz LFQPO is caused by a series of intermittent oscillations with varying frequencies. The high and low amplitude durations are $\sim3$ s and $\sim0.7$ s, respectively. These intermittent oscillations also exhibit rms-flux and frequency-flux relations. Interestingly, our findings can be explained by employing the Lense-Thirring precession interpretation \citep{Ingram2009, Ingram2011, Ingram2012a, Ingram2012b}, which is the most promising model for LFQPOs.

According to the Lense-Thirring precession model, the 4-Hz LFQPO is produced by the Lense-Thirring precession of the entire hot accretion flow around the spinning black hole. The precession occurs owing to the misalignment of angular momentum between the spinning black hole and the binary system. \citet{Ingram2009} showed that the precession frequency of the entire hot flow matches the observed frequency of the LFQPO. Because the hot flow's viscosity originates from the magnetorotational instability (MRI) \citep{Balbus1998}, all quantities of the hot flow are inherently fluctuating \citep{Krolik2002, Ingram2011}. Consequently, the stochastic turbulence caused by the MRI fluctuations stimulates the 4-Hz precession to be intermittent with the lifetime of $\sim3$ s, which is associated with the viscous timescale at the outer radius of the hot flow \citep{Ingram2009, Lachowicz2010}. The high amplitude durations can be interpreted as the durations that the hot flow experiences the steady precession, while the low amplitude durations are the moments that the 4-Hz precession become unstable and hard to be observed.

In addition, the model naturally explains the rms-flux relation by considering the propagation of the fluctuations in mass accretion rate through the hot flow \citep{Lyubarskii1997, Kotov2001, Arevalo2006, Ingram2011}. It suggests that the mass accretion rate fluctuates at different disk radii with their local viscous timescales. The fluctuations propagate inward and are coupled from large radii down to X-ray emission regions, yielding the X-ray variability on a wide range of timescales. The broad-band variability then exhibits the linear rms-flux relation because both the rms amplitude and the flux of the variation respond to the mass accretion rate fluctuations. Moreover, because the LFQPO is coupled to the broad-band variability in the Fourier power spectrum, it also exhibits the linear rms-flux relation \citep{Heil2011}.

Furthermore, the surface density of the hot flow experiences fluctuations for satisfying the mass conservation during the fluctuating propagation, thus altering the hot flow's moment of inertia and inducing the precession frequency jitter. Because the frequency fluctuations and the flux of the hot flow are both related to the mass accretion rate fluctuations, the frequency-flux relation is predicted by the Lense-Thirring precession model \citep{Ingram2011, Ingram2012b}. Finally, because the hot flow itself is communicated by bending waves \citep{Fragile2007, Ingram2011}, the timescale for restoring the coherence of its precession is faster than the viscous timescale \citep{Ingram2011, Ingram2012a}, thus explaining why the low amplitude durations of the intermittent oscillations are shorter than the high amplitude durations.

Considering the strong agreement between our HHT analysis results and the Lense-Thirring precession model, in the future we will extend our work to the remainder of the LFQPOs from XTE J1550-564. 

\section{Conclusion}\label{conclusion}
Owing to the high time-frequency resolution of HHT, we have demonstrated that the $\sim4$-Hz peak in the power spectrum of the LFQPO observed from XTE J1550-564 is broadened by a series of intermittent $\sim4$-Hz oscillations with varying frequencies, with oscillations lasting for a few seconds. We have illustrated that the HHT-derived spectrum has better time-frequency resolution than the spectrogram because the instantaneous frequency as well as the instantaneous amplitude of the oscillations can be obtained from HHT.

Because we could identify the individual oscillations by performing the HHT analysis, we discovered additional interesting characteristics of these intermittent oscillations. The rms-flux and frequency-flux relations of the oscillations confirmed the previous findings of \citet{Heil2011}. Moreover, the distributions of the high and low amplitude durations allow the lifetimes and recovery times of the oscillations to be estimated.

All of our findings can be well explained by using the Lense-Thirring precession model \citep{Ingram2009, Ingram2011, Ingram2012a, Ingram2012b}, reaffirming this model as the most promising model for LFQPOs. Given that applying the HHT analysis provides wide and deep insights into LFQPOs, we plan to extend this line of work to perform systematic HHT analysis for the remainder of the LFQPOs in XTE J1550-564 by comparing their characteristics such as the distributions of high and low amplitude durations for different LFQPOs.

\acknowledgments
We would like to thank Dr. Ming-Chya Wu for useful advice regarding the HHT analysis. We thank Professor Tomaso Belloni and Dr. Mike Nowak for helping us with modeling the QPO features by using the ISIS/SITAR package. We also thank Prof. Christopher Reynolds for useful discussion in the QPO mechanisms. The HHT MATLAB package was provided by the Research Center for Adaptive Data Analysis at the National Central University of Taiwan. The RXTE/PCA data for this research were obtained from High Energy Astrophysics Science Archive Research Center (HEASARC) online service, provided by the NASA/Goddard Space Flight Center. This research was partially supported by the Ministry of Science and Technology (MOST) of Taiwan through the grant NSC 102-2112-M-008-020-MY3. Y.H.S. especially acknowledges the support provided by the MOST through the grants NSC 102-2119-M-008 -001 and NSC 102-2112-M-008-019-MY3.

\clearpage
\bibliography{references}

\end{document}